\documentclass[prd,11pt]{revtex4-1}

\usepackage{bm}
\usepackage{amsmath}    
\usepackage{graphicx}   
\usepackage{verbatim}   
\usepackage{color}      
\usepackage{subfigure}  
\usepackage{hyperref}   
\definecolor{amaranth}{rgb}{0.9, 0.17, 0.31}
\definecolor{purple(munsell)}{rgb}{0.62, 0.0, 0.77}
\definecolor{americanrose}{rgb}{1.0, 0.01, 0.24}
\definecolor{palatinateblue}{rgb}{0.15, 0.23, 0.89}
\definecolor{royalblue(web)}{rgb}{0.25, 0.41, 0.88}
\definecolor{hanpurple}{rgb}{0.32, 0.09, 0.98}
\definecolor{beaublue}{rgb}{0.74, 0.83, 0.9}
\definecolor{carminered}{rgb}{1.0, 0.0, 0.22}
\definecolor{brightpink}{rgb}{1.0, 0.0, 0.5}
\definecolor{vividviolet}{rgb}{0.62, 0.0, 1.0}

\hypersetup{ linktoc=all,
    colorlinks, linkcolor={palatinateblue},
    citecolor={brightpink}, urlcolor={amaranth}}

\newcommand{\be}{\begin{equation}}
\newcommand{\ee}{\end{equation}}
\newcommand{\bs}{\begin{split}} 
\newcommand{\bea}{\begin{eqnarray}}
\newcommand{\eea}{\end{eqnarray}}

\begin{document}

\title{ DSR transformations with zero time delay} 

\author{Nosratollah Jafari}\email{nosrat.jafari@fai.kz}

\affiliation{Fesenkov Astrophysical Institute, 050020, Almaty, Kazakhstan;
\\
Al-Farabi Kazakh National University, Al-Farabi av. 71, 050040 Almaty, Kazakhstan;
\\
 Center for Theoretical Physics, Khazar University, 41 Mehseti Street, Baku, AZ1096, Azerbaijan}

\begin{abstract}

In doubly special relativity (DSR) theories the speed of the light usually depends on  the frequency. Thus, we expect that light from very distant stars and gamma ray bursts have additional time delays with respect to special relativity. In this paper, we will find the general DSR transformations in the first order of the Planck length which have zero time delay. 
With respect to these transformations zero time delay isn't only a consequence of special relativity, and we can find some DSR theories which behave like special relativity in the first order of the Planck length in some aspects. We notify that for a correct conclusion in the observations of the quantum gravity effects we should look to all quantum gravity effects together and not one.

\vspace{7.5cm}


\end{abstract}

\maketitle

\tableofcontents

\section{Introduction}

The well known paper by G. Amelino-Camelia:\textit{" Are we at the dawn of quantum-gravity phenomenology?}", was one of the initiating proposals for the phenomenology of the quantum gravity \cite{Amelino-Camelia:1999hpv}. The proposal for incorporating quantum effects in the special relativity transformations was the modifying the dispersion relation for the photons to     
\be    p^2=E^2 \Big[   1 + f \Big( \frac{E}{E_{QG}} \Big) \Big], \ee
  where $ E_{QG} $ is an effective quantum gravity scale, and $f$ is a model dependent function of $ E/E_{QG} $. 
 In the first order this dispersion relation will become 
  \be \label{DisRelNat}  p^2=E^2 \Big(   1 + \eta \frac{E}{E_{QG}}  \Big),   \ee
 where $\eta $   is some number. The proposal is that the speed of these photons ( for example gamma ray bursts) should be different from $c$, as given by 
 \be \label{LinearVgamma}     v=  \frac{\partial E}{ \partial p_1} = c \Big ( 1- \eta \frac{E}{E_{QG}}   \Big). \ee
 Also, time delay for a signal from a distance $D$ with energy $E$ will be 
   \be    \Delta T_{DSR} = \eta \  \frac{E}{E_{QG}}\frac{D}{c}. \ee
This time delay in special relativity is zero.

This proposal was an important motivation for introducing the doubly special relativity (DSR) 
theories. DSR theories have two invariant scales, the speed of light $c$
and the Planck energy $E_{p}=\sqrt{\hbar c^5/G}\simeq 10^{19}$
GeV. Magueijo-Smolin (MS) DSR  \cite{Magueijo:2001cr}    and Amelino-Camelia DSR  
\cite{Amelino-Camelia:2000stu, Amelino-Camelia:2002uql} are the two main examples of these theories.

In our previous paper, we have obtained the general transformations in the first order of the Planck length for DSR \cite{Jafari:2020ywd}.
These transformations and other consequent relations such as dispersion relation, rapidity, speed of the photons and time delay depend on the three parameters $\alpha_1$, $ \alpha_2 $, $\alpha_3 $. By putting these $\alpha$'s or  $ (\alpha_1 + \alpha_2 - \alpha_3 ) $ to zero we will show that  time delay can be zero as in special relativity. However, transformations do not reduce to the Lorentz transformations. Thus, we will find new transformations which are different from Lorentz transformations in the in the first order of the Planck length, but in observations they behaves like special relativity. 

We notify that testing only one parameter such as speed of the photons or time delay is not sufficient for testing quantum gravity effects. We can find some DSR theories like our DSR which we name them null DSRs that behave like special relativity in many aspects.

Zero time delay for the DSR theories has been discussed in \cite{Carmona:2022pro}. The authors has obtained the conditions for a zero time delay for the parameters in the general expressions for the generators of the infinitesimal Poincaré transformations, which are in agreement with ours. But, they didn't obtain the explicit forms of the finite DSR transformations.  These transformations can be used in searching for quantum gravity effects through time delay and speed of light tests  \cite{FermiGBMLAT:2009nfe, Vasileiou:2013vra, LHAASO:2024lub, Yang:2023kjq}. 
    
In the following section, we summarize some basics of our methods in obtaining the general DSR transformations in the first order of the Planck length  as given in \cite{Jafari:2020ywd}.  
 In section 3, we obtain the conditions for the zero time delay in our general DSR. 
 If in any DSR a  measurable parameter such as the speed of the light, the  dispersion relation, time delay or rapidity parameter will become zero we can name this DSR  as a null DSR. In section 4, we discuss these null DSR theories, and for the completeness of our discussions we also give the addition laws of the energies and the momenta.
Finally, in section 5 we have some remakes and conclusions.

\section{Modified Lorentz transformations from differential equations}
We can find the well-known Lorentz transformations in the energy momentum space 
  \bea    \left\{\begin{array}{lr}  p'_0= p_0\cosh\xi + p_1 \sinh\xi,\\
   p'_1= p_1\cosh\xi + p_0 \sinh\xi, \\
               p'_2 =  p_2, \\
               p'_3 =  p_3,
   \end{array} \right. \eea
from the second order differential equations for $p_0$ and $p_1$:
\bea  \label{seconorderlorentz}  \left\{\begin{array}{lr}  \frac{d^2p_0}{d\xi^2}- p_0=0,
       \\\\\frac{d^2p_1}{d\xi^2} -p_1=0, \end{array} \right. \eea
and the first order differential equations for the $p_2$ and $p_3$ components
 \be  \label{p2p3DiffEQLT} \left\{\begin{array}{cl}   \frac{d p_2}{d \xi}=0, \\\\
               \frac{d p_3}{d \xi} = 0.
               \end{array}\right.\ee
 The generator of the DSR theories in (3+1) dimensions is given by
              \be \label{ModGen}  N_1=  p_1 \frac{\partial  }{ \partial p_0 } + \Big(\frac{l_p}{2} \textbf{p}^2 +
              \frac{1 - e^{-2l_p p_0} }{2l_p} \Big)\frac{\partial  }{ \partial p_1 } - l_p p_1  \Big( p_j\frac{\partial }{ \partial p_j } \Big).   \ee
 In order to include a wider range of theories, we modify this generator by generalizing to preserve parity and time-reversal symmetry: 
\be \label{3DimGenerator} \Tilde{N}_1=   ( p_1 +  l_p\beta_0 p_0 p_1 ) \frac{\partial  }{ \partial p_0 } + 
 \Big( p_0 +  l_p\beta_1 p_0^2+ l_p\beta_2 \textbf{p}^2 \Big)\frac{\partial  }{ \partial p_1 } + 
 l_p\beta_3 p_1 \Big( p_j \frac{\partial  }{ \partial p_j } \Big)  + l_p\beta_4 \epsilon_{1jk} p_j \frac{\partial  }{ \partial p_k },\ee
 where $\beta_0$, ..., $\beta_4$ are arbitrary real numbers. 

From this modified generator we can obtain modified second order differential equations,
 \bea \label{SecOrderDEBoost}  \left\{\begin{array}{lr}  \frac{d^2p_0}{d\xi^2}= p_0  + l_p a_0 p_0^2  + l_p a_1 (\frac{dp_0}{d\xi})^2 +
     \beta_2 (p_2^2 +p_3^2),
  \\\\
  \frac{d^2p_1}{d\xi^2} =   p_1 + l_p a_3 p_1\frac{dp_1}{d\xi}. \end{array} \right. \eea
For the $ p_2$ and $p_3$ components the modified differential equations are
\be  \label{p2p3_DifEq} \left\{\begin{array}{cl}  
    \frac{d p_2}{d \xi}  &=  l_p\beta_3 p_1p_2 - l_p\beta_4 p_0p_3, \\\\
    \frac{d p_3}{d \xi}  &=  l_p\beta_3 p_1p_3 + l_p\beta_4 p_0p_2.  \end{array}\right.\ee

From Eq.~(\ref{SecOrderDEBoost}) and Eq.~(\ref{p2p3_DifEq}) we can obtain the finite-boost transformations to the first order in $l_p$ as
\bea   \label{3DimTransform} \left\{\begin{array}{lr}  p'_0= p_0\cosh\xi + p_1 \sinh\xi 
    \\ \hspace{ 1cm}+ l_p( \alpha_1 p_0^2 + \alpha_2 p_1^2) \sinh^2 \xi + l_p( \alpha_2 p_0^2 + \alpha_1 p_1^2) \cosh^2 \xi 
    \\ \hspace{ 1cm}+ 2 l_p(\alpha_1+ \alpha_2) p_0 p_1 \sinh\xi \cosh \xi - l_p(\alpha_2 p_0^2 + \alpha_1 p_1^2)\cosh \xi
    -  l_p\alpha_3 p_0 p_1 \sinh\xi  \\~~~~~~~ + l_p\beta_2 (p_2^2 + p_3^2 )(\cosh\xi -1),\\\\
   p'_1= p_1\cosh\xi + p_0 \sinh\xi 
   \\ \hspace{1cm}  + l_p \alpha_3  p_0 p_1 \sinh^2 \xi  + l_p \alpha_3  p_0 p_1\cosh^2 \xi
   \\ \hspace{1cm}  + l_p \alpha_3  (p_0^2 +p_1^2)\sinh\xi \cosh \xi   -l_p \alpha_3  p_0 p_1 \cosh \xi - l_p(\alpha_2 p_0^2 +
   \alpha_1 p_1^2)\sinh\xi  \\ ~~~~~~ + l_p\beta_2 (p_2^2 + p_3^2 )\sinh\xi   ,\\\\ 
   p'_2= p_2 + l_p (\beta_3 p_1 p_2 -\beta_4 p_0 p_3 )\sinh\xi + \l_p (\beta_3 p_0 p_2 -\beta_4 p_1 p_3 )(\cosh\xi-1),\\\\
     p'_3= p_3 + l_p (\beta_3 p_1 p_3 +\beta_4 p_0 p_2 )\sinh\xi + \l_p (\beta_3 p_0 p_3 +\beta_4 p_1 p_2 )(\cosh\xi-1), \end{array} \right. \eea
where
         \be  \label{AlphaParameters}
    \alpha_ 1\equiv \frac{\beta_0 + 2\beta_1 -\beta_2 -\beta_3 }{3}, ~~~ 
   \alpha_ 2\equiv \frac{\beta_0 -\beta_1 +2\beta_2 + 2\beta_3 }{3},\; \textrm{and}~~~ \alpha_ 3\equiv \frac{\beta_0 +2\beta_1+2\beta_2+ 2\beta_3 }{3},
   \ee
   for convenience. These transformations are finite-boost DSR transformations to first order in $l_p$ but for all orders of rapidity $\xi$.

\section{DSR with Zero time delay}

Speed of light in DSR Eq.~(\ref{3DimTransform}) differs from c and is given by 
 \be \label{LinearVelocity}    v=  \frac{\partial E}{ \partial p_1} \simeq  c \Big[1 + (\alpha_1 + \alpha_2 - \alpha_3  ) l_p E \Big], \ee  
 If we put 
 \be \label{ZeroTimCond} \alpha_1 + \alpha_2 - \alpha_3 =0, \ee
 the speed of light  will be the same as $c$
 \be  v=c.   \ee 
For the time delay we have
  \be  \label{TimeDelayDSR} \Delta t_{DSR}  =  (\alpha_1 + \alpha_2 - \alpha_3 ) \frac{ l_p E D}{c}. \ee
  If we put the condition  Eq.~(\ref{ZeroTimCond}) then the time delay will become zero
 \be  \label{TimeDelayDSR} \Delta t_{DSR}  =  0, \ee
 as in special relativity.
 
 All of the transformations in Eq.~(\ref{3DimTransform})  with the condition Eq.~(\ref{ZeroTimCond}) have zero time delay. This condition in terms of the $\beta_i$ will be $ \alpha_1=\beta_1,~ \alpha_2=\beta_0- \beta_1,~  \alpha_3= \beta_0$.

\section{Null DSR }

If in any DSR a measurable parameter such as the speed of the light, the  dispersion relation, time delay or rapidity parameter will become zero we can name this DSR  as a null DSR.

  Dispersion relation for the transformations Eq.~(\ref{3DimTransform}) in the first order of $l_p$  will be
  \be  \label{DispRelDSR} p_0^2- \textbf{p}^2 - 2 l_p \alpha_2 p_0^3 + 2 l_p ( \alpha_3 - \alpha_1 )p_0 \textbf{p}^2 =m^2. \ee
Here, $m$ is the rest mass of the particle. For the special relativistic dispersion relation the conditions are
\be    \label{CondAlpha} \alpha_1 = \alpha_2= \alpha_3 =0,  \ee these conditions in terms of $ \beta_\mu$ will be $ \beta_0=\beta_1=0 $, and $ \beta_2=-\beta_3$. Putting these conditions in Eq.~(\ref{DispRelDSR}), the dispersion relation will become the special relativistic dispersion relation 
\be  \label{DisRelSR} p_0^2- \textbf{p}^2 =m^2. \ee
 Also, the expressions for  $\cosh\xi$ and $\sinh\xi$ which are
 \be  \label{CosSinDSR} \cosh\xi= \frac{p_0- l_p \alpha_1 \textbf{p}^2 - l_p \alpha_2 p_0}{m} = \frac{p_0}{m},~~~
           \sinh\xi= \frac{ |\textbf{p}|- l_p  \alpha_3  p_0 |\textbf{p}|}{m}=\frac{ |\textbf{p}|}{m}.\ee
 will take the  special relativistic value
\be   \label{CosSinSR}  \cosh\xi = \frac{p_0}{m},~~~      \sinh\xi=\frac{ |\textbf{p}|}{m}.\ee

 By putting the values of $ \alpha_1$ to $\alpha_3 $ parameters in Eq.~(\ref{3DimTransform}), we find the following transformations
        \bea   \label{SRTransform} \left\{\begin{array}{lr}  p'_0= p_0\cosh\xi + p_1 \sinh\xi +
        l_p\beta_2 (p_2^2 + p_3^2 )(\cosh\xi -1),\\\\
   p'_1= p_1\cosh\xi + p_0 \sinh\xi + l_p\beta_2 (p_2^2 + p_3^2 )\sinh\xi   ,\\\\ 
   p'_2= p_2 - l_p (\beta_2 p_1 p_2 + \beta_4 p_0 p_3 )\sinh\xi - \l_p (\beta_2 p_0 p_2 +\beta_4 p_1 p_3 )(\cosh\xi-1),\\\\
     p'_3= p_3 + l_p (\beta_4 p_0 p_2 -\beta_2 p_1 p_3 )\sinh\xi + \l_p (\beta_4 p_1 p_2 - \beta_2 p_0 p_3 )(\cosh\xi-1), \end{array} \right. \eea
  These transformations are the finite-boost DSR transformations in the first order of Planck length $l_p$, with an unmodified special relativistic dispersion relation, and they differ from usual Lorentz transformations. We have also two other free parameters $\beta_2$ and $\beta_4$.  All of the aspects of these DSR Eq.~(\ref{SRTransform}) are the same as specail relativity expect to the addition laws of the energies and the momenta. 

  \subsection{ Addition laws for the energies and the momenta}

  For completing our discussion we give the addition laws for the energies and momenta for our transformations in the Eq.~(\ref{3DimTransform}) \cite{Wang:2013bta}.
  By introducing the new parameters 

  \be \label{newParam}
  \delta_1=-2(\beta_0 + \beta_1 + 2\beta_2 ),~~~  
  \delta_2=2\beta_2, ~~~ 
  \eta_1=\eta_2= -(\beta_0 + 2\beta_1 + 2\beta_2 ), \ee

   addition laws will be
  \bea   \label{AdditionLawDSR} \left\{\begin{array}{lr}  E_a \oplus E_b=E_a+E_b + l_p \delta_1 E_a E_b +l_p \delta_2 E_a E_b
  \\\\\
  \textbf{p}_a \oplus \textbf{p}_b= \textbf{p}_a+ \textbf{p}_b + l_p \eta_1 E_a \textbf{p}_b +l_p \eta_2 E_b \textbf{p}_a  \end{array} \right. \eea
  where $\oplus $ is additive sign for addition in the DSR.

If we take $ \delta_1, ~\delta_2, ~ \eta_1~ ,\eta_2 $ to zero  then the addition laws in Eq.~(\ref{AdditionLawDSR})  will become additive as in special relativity. Thus, all of the of $ \beta_\mu$ will become zero and transformations will reduce to the usual Lorentz transformations.

\section{Conclusion and some remarks}

We have obtained the transformations in the first order of the Planck length which have zero time delay. Beside to this we found a null DSR theoey, in which all  of the measurable parameter such as the speed of the light, the  dispersion relation, time delay or rapidity parameter except to the addition of the energy momentum variables are zero.  This DSR has been obtained by putting all $\alpha_i$ to zero. However, we can obtain other constraints on the $\beta_\mu$ parameters from astrophysical and cosmological observations. But, frequently in these observations we reach to the null results. In other words, we have encounter a null DSR result.

\section{ Acknowledgment}
This research is funded by the Science Committee of the Ministry of Science and Higher Education of the Republic of Kazakhstan Program No. BR21881880.

\bibliography{main}

\end{document}